\documentclass[aps,prb,twocolumn,showpacs,amsmath,amssymb,superscriptaddress]{revtex4}
\usepackage{graphicx}
\usepackage{dcolumn}
\usepackage{epstopdf}
\usepackage{bm}
\usepackage{color}

\begin{document}
\title{Adsorption of diatomic halogen molecules on graphene: A van der Waals density functional study}

\author{A.~N. Rudenko}
\email[]{rudenko@tu-harburg.de}
\author{F.~J. Keil}
\affiliation{Institute of Chemical Reaction Engineering, Hamburg University of Technology, Eissendorfer Str.~38, 21073 Hamburg, Germany}
\author{M.~I. Katsnelson}
\affiliation{Radboud University Nijmegen, Institute for Molecules and Materials, Heijendaalseweg 135, 6525 AJ Nijmegen, The Netherlands}
\author{A.~I. Lichtenstein}
\affiliation{Institute of Theoretical Physics, University of Hamburg,
Jungiusstrasse 9, 20355 Hamburg, Germany}
\date{\today}

\begin{abstract}
The adsorption of fluorine, chlorine, bromine, and
iodine diatomic molecules on graphene has been investigated using 
density functional
theory with taking into account nonlocal correlation effects by
means of vdW-DF approach. It is shown that the van der Waals
interaction plays a crucial role in the formation of chemical
bonding between graphene and halogen molecules, and is therefore
important for a proper description of adsorption in this system.
In-plane orientation of the molecules has been found to be more stable than
the orientation perpendicular to the graphene layer. In the cases
of F$_2$, Br$_2$ and I$_2$ we also found an ionic contribution to
the binding energy, slowly vanishing with distance.
Analysis of the electronic structure shows that ionic interaction
arises due to the charge transfer from graphene to the molecules.
Furthermore, we found that the increase of impurity concentration
leads to the conduction band formation in graphene due to
interaction between halogen molecules. In addition, graphite
intercalation by halogen molecules has been investigated. In the
presence of halogen molecules the binding between graphite layers
becomes significantly weaker, which is in accordance with the results of
recent experiments on sonochemical exfoliation of intercalated
graphite.
\end{abstract}

\maketitle

\section{Introduction}

A monolayer of graphite, commonly known as graphene, the first
truly two-dimensional crystal (one-atom-thick), which became
experimentally available in the last years.\cite{Novoselov} Today,
graphene is at the focus of many research activities worldwide.
Remarkable electronic properties of graphene make this material a
promising candidate for a large variety of electronic
applications.\cite{Geim,Graphene-RMP}

There have been a number of publications focused on theoretical
investigation of covalent-, ionic- and metal-bonded impurities on
graphene (for example, see
Refs.\onlinecite{Wehling-PRB,Wehling-APL,Khomyakov,Johll}). These
studies are primarily based on the density functional theory (DFT)
combined with standard functionals for electronic
exchange-correlation effects, such as local density approximation
(LDA) and generalized gradient approximation (GGA). The employment
of (semi-)local approximations looks quite reasonable so far as
the systems with either covalent or metallic bonding are
concerned. However, in the case of weakly bounded systems, such as
molecules or molecular compounds, dispersion forces play a
significant role. Such interactions are essentially nonlocal and
therefore cannot be properly taken into account using local
or semi-local density functionals.

In this work, we will focus on molecular impurities on graphene. In
particular, we consider diatomic halogen molecules on an ideal
graphene lattice. The motivation for this choice is the following.
First, these molecules themselves are covalently bonded and
represent an example of realistic weakly reactive adsorbates.
Second, because of the relative chemical inertness of the
molecules, the interaction between these molecules and the graphene
sheet is rather weak and cannot be properly described within the
electronic structure methods widely used nowadays in condensed
matter physics (such as conventional LDA or GGA approximations in
the density functional theories). Accurate calculations of the
adsorption energies, which are necessary to judge how the
impurities can be removed from the surface, are beyond these
standard approaches which make their calculation a challenging
problem. Last but not least, among the fundamental aspects of studying
impurities on graphene, in some cases molecularly doped graphene
represents itself an unavoidable outcome of its production
processes. Recently Windekvist \emph{et al.}\cite{Br-intercalat}
have reported a method of graphene fabrication based on
sonochemical exfoliation of bromine-intercalated graphite. It was
shown that ultrasonic treatment of graphite with absorbed bromine
molecules results in intensive graphene-flakes formation. It is
worth mentioning that intercalated graphite by itself has many
potential applications.\cite{intercalation-book}

In this paper we show that the adsorption of halogens on graphene
is determined generally by the van der Waals (vdW) interactions. 
Besides the vdW interaction we also found a
considerable ionic contribution for F$_2$ and I$_2$ adsorbates. We
demonstrate that the ionic component of the interaction arises due
to the electronic transfer from graphene to the molecules.
Furthermore, the ionic interaction gives rise to a long-range tail
of adsorption curves, which determines the adsorption at large
distances.

In this work we also apply a van der Waals density functional
approach (vdW-DF)\cite{Dion,Thonhauser} to investigate structural and
energetic properties of graphite intercalated by different
diatomic halogen molecules. We show that the interlayer distance
and binding energies in graphite are also dependent on the
exchange-correlation functional. Using vdW-DF method we have
calculated interplanar binding of graphite layers in the presence
of halogen molecules, which was found to be significantly weaker
than the binding in pure graphite.

The rest of the paper is organized as follows. In Sec. II we
briefly describe methods of the calculations. In Sec. III we
present the results of adsorption of F$_2$, Cl$_2$, Br$_2$, and
I$_2$ on graphene. Section IV is devoted to the analysis of
electronic structure and ionic interaction between graphene and
its adsorbates. The results of graphite intercalation by halogen
molecules and corresponding discussion are given in Sec. V. In
Sec. VI we briefly summarize our results.

\section{Computational details}

\subsection{Crystal structure}

The structure of the systems under consideration was modelled
using a supercell approach. The supercell consists of one graphene
layer with a molecule on the top and a vacuum region to avoid
spurious interaction between periodic images of the supercell in
$z$-direction. In order to examine the effects of impurity
concentration two supercells containing 24 and 32 carbon atoms
per single molecule were considered. The corresponding graphene
layers were constructed from rectangular $(3 \times 2\sqrt 3)a$
and $(4 \times 2\sqrt 3)a$ unit cells, respectively. The height of
the supercells perpendicular to the surface was chosen to be 40
\AA. Investigating the supercells of such a size allows us to judge how
the interactions between molecules can affect the system
properties.

We used the lattice constant of graphene equal to $a$=2.459 \AA \
in accordance with experimentally obtained value for graphite at
low temperatures.\cite{Baskin} The distortion of graphene layer in
the presence of adsorbates was not taken into account. Since the
interaction between the molecules and graphene is pretty weak the
distortions of carbon lattice is supposed to be small, anyway.
Interatomic distances in the halogen molecules were also taken
from the experimental data.\cite{Pauling}

\begin{figure}[!ht]
\includegraphics[width=0.48\textwidth, angle=0]{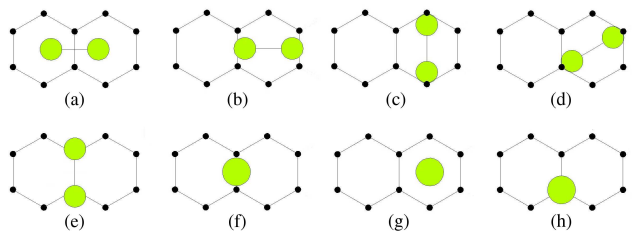}
\caption{Schematic representation of different high-symmetry adsorption sites
for diatomic molecules on graphene surface (top view). In the case of in-plane
orientation of the molecules the center of mass is located above: 
(a) a bridge
site (the middle of C-C bond) along \emph{x}-axis, (b) a hollow site along
\emph{x}-axis, (c) a hollow site along \emph{y}-axis, (d) a hollow site along
$\frac{\sqrt{3}}{2}x$-axis, (e) a bridge site along \emph{y}-axis. Molecules
perpendicular to the surface are located above: (f) a bridge site,
(g) a hollow site, (h) a top site.}
\label{structures}
\end{figure}

Both in-plane and perpendicular-to-plane orientations of the
molecules on graphene have been investigated. We considered five
in-plane and three perpendicular-to-plane configurations
corresponding to high-symmetry adsorption sites as illustrated in
Fig.\ref{structures}.

\subsection{DFT computational details}

Ground-state energies and electronic density
distributions have been calculated using the plane-wave pseudopotential 
method as implemented in the {\sc Quantum-ESPRESSO} simulation
package.\cite{espresso,pseudo} Exchange and correlation effects
have been generally taken into account using vdW-DF nonlocal
functional (see the next subsection for details). In order to
obtain some comparative results the GGA functional in
Pedrew-Burke-Ernzerhof (PBE) parametrization\cite{pbe} also was
used.

In our calculations we employed an energy cutoff of 30 Ry for the
plane-wave basis and 300 Ry for the charge density.
Self-consistent calculations of the Kohn-Sham equations were carried out
employing the convergence criterion of 10$^{-8}$ Ry. For accurate
Brillouin-zone integration the tetrahedron
scheme\cite{tetrahedron} and (16 $\times$ 16 $\times$ 1)
Monkhorst-Pack {\bf k}-point mesh \cite{monkhorst} were used.

Geometry optimization algorithms were not used in our study. Atoms
in graphene layer and interatomic distances in the molecules were
assumed to be fixed. In order to find equilibrium distances of
adsorbates we performed a set of explicit energy calculations for
different separations from the surface. The most stable positions
of the molecules were determined comparing the energies of
different high-symmetry adsorption sites.

\subsection{Non-local corrections}

In the recent years several reasonable methods have been developed
to improve the performance of DFT in description of sparse
matter.\cite{Andersson,Silvestrelli-JPCA,Nguyen} Most of them are
based on the adiabatic connection fluctuation-dissipation framework
(ACFD), which is a computationally very demanding approach
and, in practice, is limited to small systems only.\cite{Niquet,Harl}

Here, to calculate adsorption energies, we use an approach
proposed by Dion \emph{et al}.\cite{Dion} The main idea of the
method is that the ACFD correlation energy can be significantly
simplified by applying a series of reasonable approximations
resulting in the expression dependent on the electronic density
only. Although being developed only recently, up to now this
method has been widely verified on a large variety of examples and
shows the transferability across a broad spectrum of interactions,
such as ionic, covalent, and vdW.\cite{Langreth}

\begin{figure*}[bpt]
\includegraphics[width=0.80\textwidth, angle=0]{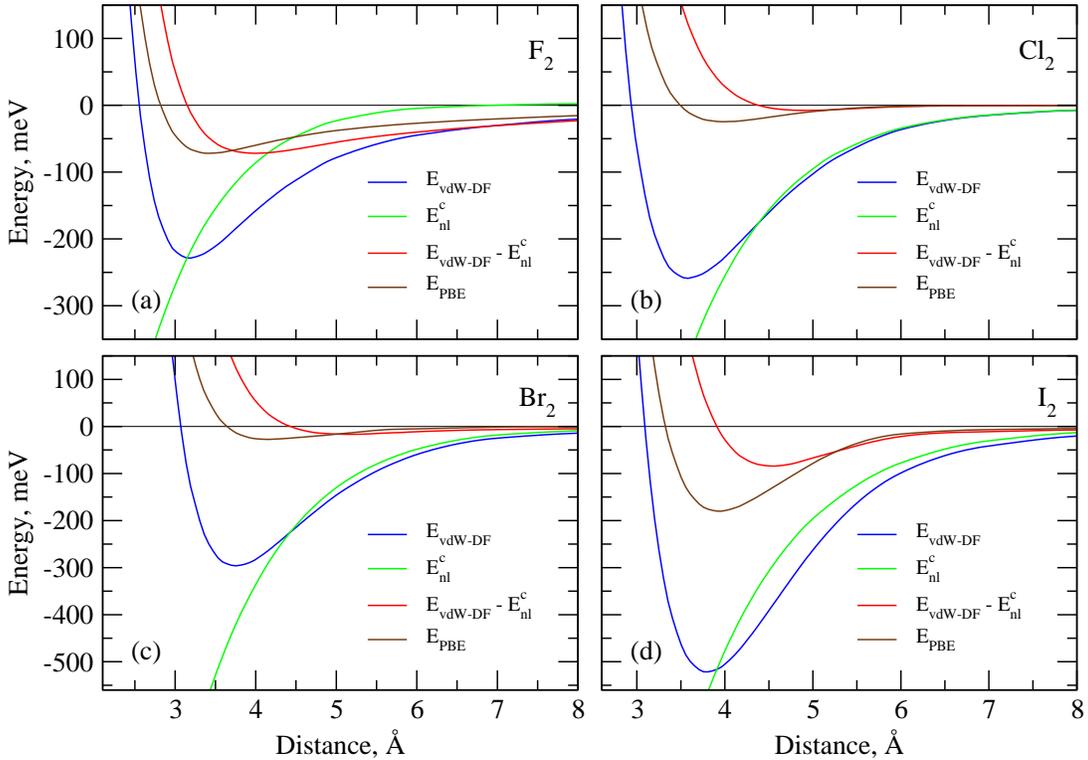}
\caption{Adsorption energies of halogen molecules on graphene as a function of the separation between graphene and the center of mass of
the molecules: (a) F$_2$, (b) Cl$_2$, (c) Br$_2$, (d) I$_2$. Blue, green and red curves correspond to the total binding energy calculated
within vdW-DF approach, nonlocal vdW correction (Eq.\ref{nonlocal}), and to the binding energies excluding vdW correction, respectively.
Brown curve corresponds to the PBE binding energy. Adsorption curves are shown only for the most stable molecular configuration
(Fig.\ref{structures}a).
}
\label{ad_curves}
\end{figure*}

In the vdW-DF method, the exchange-correlation energy functional
consists of several contributions,
\begin{equation}
E_{xc}[n]=E_x^{revPBE}[n]+E_c^{LDA}[n]+E_c^{nl}[n],
\label{exc}
\end{equation}
where the first term corresponds to the exchange part of the revised
PBE (revPBE) functional,\cite{Zhang-Yang} $E_c^{LDA}$ is the
LDA correlation energy, and $E_c^{nl}[n]$ is the nonlocal
correlation correction, which is calculated in the following way:
\begin{equation}
E_c^{nl}=\frac{1}{2}\int d^3r d^3r' n(r)\phi(r,r')n(r'),
\label{nonlocal}
\end{equation}
where $n(r)$ is the electronic density and $\phi(r,r')$ is a
function incorporating many-body density response (for details see
Ref.\onlinecite{Dion}). It should be emphasized that in our work
we evaluate the nonlocal correction (Eq.\ref{nonlocal}) in a
perturbative way, i.e. using only GGA-based (semi-local)
electronic density distribution. This choice
looks quite reasonable since it has been previously shown that the
effects due to the lack of self-consistency are negligible.\cite{Thonhauser}

\section{Adsorption of molecules on graphene}

As we have already mentioned, we consider two different sizes of the
supercell in order to figure out the influence of intermolecular
binding on the investigated properties. We found that the adsorption
energy at higher concentrations of halogens is slightly lower,
however, this difference does not exceed 10 meV and could be
considered as insignificant. Due to this fact we do not expect any
considerable changes in adsorption for different molecular
concentrations. In the rest of this section we present the results
only for the smaller supercell containing 24 carbon atoms.

We found the in-plane bridge position (Fig.\ref{structures}a) to
be the most stable configuration for all the molecules under
consideration. The binding energies for this particular geometry
are shown in Fig.\ref{ad_curves} as a function of the distance
between the surface and the center of mass of the molecule.

It is clear from Fig.\ref{ad_curves} that the account of the vdW
interaction gives a considerable contribution to the binding
energy for each molecule examined. One can see that the larger the
radii of atoms in the molecule the stronger is the vdW binding to the
surface. This fact can be easily figured out using the
interpretation of the vdW interaction as an interaction of induced
dipoles.

One can see from Fig.\ref{ad_curves} that the binding energy of
the molecules decreases quite slowly with the distance. This fact
is the most evident for the fluorine molecule
(Fig.\ref{ad_curves}a), due to the relatively small energy scale.
Moreover, the behavior of fluorine binding with graphene at large
distances is fairly close to inverse proportionality and, therefore,
indicates the presence of a strong ionic interaction. Indeed, at
the distance of several angstroms the electronic density of
graphene does not overlap with that of adsorbed molecules, which
means that covalent interaction cannot occur in this case.
Furthermore, the slowly vanishing tail of adsorption curves remains,
even without taking into account the vdW correction, which
confirms the ionic nature of such long-range interaction.

    \begin{table*}[!bt]
    \centering
    \caption[]{Binding energies in graphene-halogen system for different high-symmetry adsorption sites of the molecules, and corresponding equilibrium distances (in parentheses). Adsorption sites denoted as in Fig.\ref{structures}. Values are given in meV and \AA, respectively.}
    \label{values}
 \begin{tabular}{ccccccccccc}
      \hline
      \hline
 &  \multicolumn{5}{c}{in-plane} & & \multicolumn{3}{c}{perpendicular-to-plane} \\
\cline{2-6}
\cline{8-10}
         & a       & b       & c       & d       & e       &  & f      & g      & h       \\
     \hline
 F$_2$   & -231 & -227 & -225 & -226 & -222 &  & -214 & -187 & -218  \\
         & (3.17)  & (3.23)  & (3.24)  & (3.23)  & (3.26)  &  & (3.47)  & (3.70)  & (3.46)  \\
 Cl$_2$  & -259 & -248 & -249 & -250 & -251 &  & -208 & -193 & -210  \\
         & (3.58)  & (3.65)  & (3.64)  & (3.65)  & (3.63)  &  & (4.21)  & (4.34)  & (4.21) \\
 Br$_2$  & -296 & -289 & -287 & -285 & -288 &  & -229 & -207 & -214  \\
         & (3.74)  & (3.77)  & (3.77)  & (3.77)  & (3.78)  &  & (4.51)  & (4.67)  & (4.45)  \\
 I$_2$   & -523 & -514 & -508 & -501 & -512 &  & -375 & -352 & -379 \\
         & (3.80)  & (3.84)  & (3.83)  & (3.83)  & (3.81)  &  & (4.74)  & (4.85)  & (4.73)  \\
      \hline
      \hline
    \end{tabular}
    \end{table*}

In the case of originally neutral constituents of the system, the
ionic interaction could emerge only due to the charge transfer
between them. Comparison of the binding energy for different halogens
at relatively large distances (more than 6 \AA) shows that
in the absence of the vdW interaction the binding of graphene with
fluorine is stronger than with other molecules. Assuming that this
interaction is essentially ionic, we expect the larger charge
transfer in F$_2$ case. By the same reasoning the charge
imbalance for I$_2$ should be larger than for both Br$_2$ and
Cl$_2$. A more detailed discussion on the ionic interaction will be given in 
the next section.

For comparison we also show in Fig.\ref{ad_curves} the results
obtained by the standard semi-local GGA functional in PBE
parametrization. While the equilibrium distances of vdW-DF and PBE
approaches agree reasonably well, the adsorption energies in the
case of PBE are highly overestimated. The latter fact is not
surprising since GGA-like functionals do not incorporate nonlocal
correlation effects, which are responsible for the vdW
interaction. It should be also pointed out that PBE functional
yields the wrong asymptote for the adsorption curves with the
exception of fluorine, where the vdW contribution at the large
distances is not as large as the ionic one.

Adsorption energies for different orientations of the molecules are 
summarized in Table \ref{values}. For both in-plane and
perpendicular-to-plane orientations the less stable configurations
correspond to the hollow adsorption site
(Fig.\ref{structures}b,c,d,g). This result is quite reasonable as
long as the vdW interaction dominates in the system. Indeed, at
the same distance the strength of the vdW interaction is
determined by electronic densities corresponding to interacting
components of the system.\cite{Andersson} It turns out that the
hollow site corresponds to the minimum of electronic density in
graphene, which accounts for the lower vdW interaction in this
case.

The difference between various adsorption energies of in-plane 
(or perpendicular-to-plane) orientations of the molecules is small. 
This is especially clear for the rotated molecules at the hollow
adsorption site (Fig.\ref{structures}b,c,d). In contrast, the
adsorption energies of in-plane orientations are sufficiently different from 
the energies of perpendicular-to-plane orientations, exceeding thermal 
energy at room temperature ($\sim$25 meV).

We assume that the molecular bond length cannot be significantly
changed during the adsorption process because the obtained
molecule-surface interaction is much weaker than the
intramolecular binding in halogens.\cite{Pauling} Due to the same
reason we do not expect the dissociative adsorption of halogens on
graphene, which usually occurs for light molecules on metallic
surfaces.\cite{gross} We note, however, that a careful description
of this phenomenon requires atomic dynamics to be taken into
account, and therefore should be investigated separately.

\section{Electronic structure and ionic interaction}

In the previous section we have shown that the vdW interaction
gives a significant contribution to the adsorption energy, and,
therefore, plays an important role in formation of bonding between
halogen molecules and graphene. However, besides the vdW part
there is another source of interaction which gives rise to the
bonding. In particular, this contribution especially important for 
fluorine and iodine adsorbates. In order to clarify the nature of such
an interaction let us analyze the density of electronic states
in the absence of the vdW forces.

\begin{figure}[!tbp]
    \vspace{0.5cm}
\includegraphics[width=0.44\textwidth, angle=0]{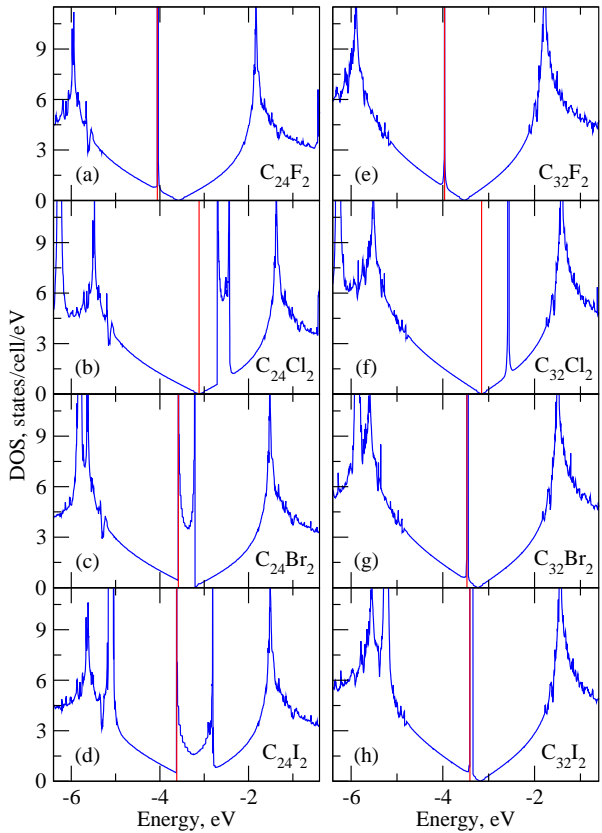}
\caption{Total electronic density of states of halogen molecules
adsorbed on graphene calculated for different supercells. Left
(right) panel corresponds to the supercell containing 24 (32)
carbon atoms. Vertical line accentuates the Fermi energy.
Densities of states are shown for the respective ground state
distances obtained without vdW corrections (see
Fig.\ref{ad_curves}).} \label{dos}
\end{figure}

In Fig.\ref{dos} we have plotted the density of states (DOS) for
each considered molecule in two different supercells, which is
calculated for the lowest-energy structural state without taking
into account nonlocal correction to correlation energy (excluding
third term in Eq.\ref{exc}). In the vicinity of the Fermi level
DOS exhibits resonances which are entirely localized on
$p$-orbitals of halogen molecules. The main differences between the
spectra for different molecular concentrations are related to the width
of impurity bands. In both cases the impurity states close to the
Fermi level remain unoccupied.

Analysis of the electronic bands shows that the impurity states
close to the Fermi level correspond only to one electron per spin,
i.e. to one electronic level in non spin-polarized case. This
means that the observed impurity band at high concentrations of molecules 
cannot be associated with the hybridization between molecular and
graphene orbitals. In fact, broadening of the impurity level
caused by the hybridization between molecular $p_x$-orbitals from
different images of the supercell forming parabolic-like DOS,
which is very similar to that of a one-dimensional chain. Indeed,
for the most energetically favorable configuration
(Fig.\ref{structures}a) the distance between molecules in
$x$-direction of the supercell is smaller than the distance in
other directions. As the distance between atoms in the molecule
increases, the separation between different molecules becomes
smaller, which leads to hybridization of molecular orbitals and
broadening of the electronic levels. This also accounts for the
absence of broad impurity bands for lower molecular concentrations where the
lateral interactions between molecules are negligible. The shape
of the band for the smallest supercell (24 carbon atoms) clearly
shows formation of {\it one-dimensional} band, with typical
divergencies (van Hove singularities) at the edges. The width of
this band, that is, the distance between the singularities, is
estimated as twice the effective hopping parameter.

For low concentrations, the impurity states are located
well below or above the conical (Dirac) point, which is easily
recognizable in the density of states. However, this is not the
case for Br$_2$ and I$_2$ adsorbed at their highest concentrations
(Fig.\ref{dos}c,d). In this situation, the Dirac point lies inside
the impurity band. Moreover, since the impurity states for Br$_2$
and I$_2$ are broadened enough, they can be considered as
conduction bands leading to a metallic behavior of the system.
Therefore, the conductivity of graphene can be adjusted varying
the concentration of Br$_2$ and I$_2$ adsorbates. There are preliminary
experimental evidences that doping by iodine can essentially increase the
conductivity of graphene.\cite{grigorieva}

As can be seen from Fig.\ref{dos} the Fermi level is shifted below
the Dirac point for the all molecules except chlorine. A shift
downward means that the electrons are donated by graphene to
F$_2$, Br$_2$ and I$_2$ molecules (acceptors). As discussed above,
the electronic transfer is not surprising for the investigated
systems. It turns out that the assumption of an ionic interaction
between graphene and its adsorbates, given in the previous
section, is in a good agreement with the peculiarities of
electronic structure.

\begin{figure}[!bpt]
    \vspace{0.5cm}
\includegraphics[width=0.47\textwidth, angle=0]{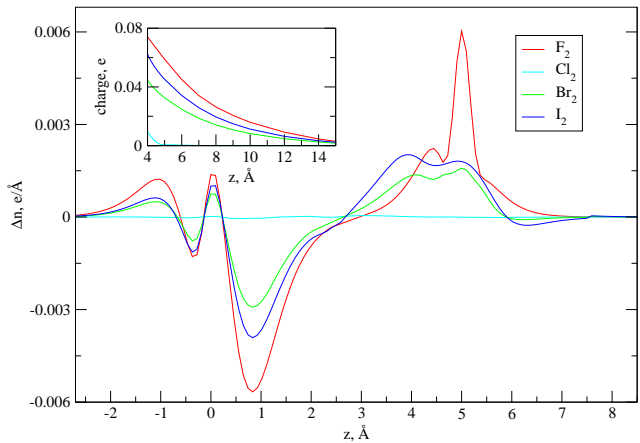}
\caption{In-plane-averaged electron density difference calculated
along $z$-axis of the supercell for F$_2$, Cl$_2$, Br$_2$, and
I$_2$ molecules on graphene. $z=0$ corresponds to position of the
carbon layer. In the inset an integrated amount of the charge
transferred from the surface to the molecules as a function of the
distance is displayed.} \label{densities}
\end{figure}

For a more detailed description of the ionic interaction, let us
analyze the charge transfer phenomenon. As the molecules come
closer to graphene, redistribution of the electronic density
results in the formation of an interface dipole layer, which is
responsible for the ionic interaction in the system. In
Fig.\ref{densities} we show the plane-averaged density differences
$\Delta n(z)$ calculated along $z$-axis of the supercell for each
investigating molecule at the distance of 5 \AA \ from the
surface. The most noticeable feature of Fig.\ref{densities} is
that there are almost no changes in electron density for chlorine
in comparison with other molecules. This seems to be a reasonable
result because the Fermi energy coincides in this case with the
Dirac point. The electron redistribution $\Delta n(z)$ for the
other molecules has nearly the same shape, which is fairly close
to the shape of a typical dipolar distribution.  $\Delta n(z)$
exhibits a main minimum (maximum) corresponding to localization of
the charge near the carbon layer (molecule). In the inset of
Fig.\ref{densities} we also show an integrated charge, transferred
from graphene to the molecules as a function of the distance
between them.\cite{comment} As expected, the amount of the
transferred charge decreases as the distance between graphene and
the molecules increases. As can be seen from Fig.\ref{densities},
at the equilibrium distance fluorine molecule accepts more
electrons than either I$_2$ or Br$_2$. This fact completely agrees
with our previous results and explains the large ionic interaction
between fluorine molecule and graphene.

\section{Intercalation of graphite}

It is known that the van der Waals interaction plays an important
role in understanding and proper description of the interplanar
bonding in graphite.\cite{Grt-vdw1,Grt-vdw2,Grt-vdw3} In spite of
the fact that the standard local density approximation (LDA)
completely misses dispersion interaction (at least for long
distances), it reproduces correctly the interlayer spacing in
graphite and even the binding energy in some cases.\cite{Grt-vdw2}
This surprising result seems, however, to be accidental because it
is impossible to reproduce accurately the experimental
compressibility of graphitic structures within the LDA.\cite{PCCP}
Recently it was shown\cite{chakarova} that the vdW-DF approach
yields not only the proper distance between graphite sheets, but
also the binding energy, which is very close to the experimental
estimations available today.\cite{benedict,zacharia} The vdW-DF
method was also successfully applied to the investigation of
energetics in potassium-intercalated graphite yielding a good
agreement with experiment.\cite{Ziambaras}

In this section we use the vdW-DF approach to describe structural
and energetic properties of graphite intercalated by halogen
molecules. We consider graphite in \emph{AB}-stacking containing
48 carbon atoms and 2 halogen molecules per unit cell.
Absorbed molecules are assumed to be uniformly distributed
over graphite occupying in-plane bridge sites in accordance with
the most stable position on graphene layer
(Fig.\ref{structures}a). Other parameters of the calculations were
taken to be the same as described in Sec.IIB.

    \begin{table}[!bt]
    \centering
    \caption[Bset]{Equilibrium interlayer distances, absorption energies, and interlayer binding energies for the pure and
halogen-intercalated graphite. Absorption and binding energies are given in meV per absorbed molecule and in meV per carbon atoms in the
supercell, respectively.}
    \label{grt_table}
\begin{ruledtabular}
 \begin{tabular}{ccccccc}
                          & C$_{24}$ & C$_{24}$F$_2$ & C$_{24}$Cl$_2$ &  C$_{24}$Br$_2$ & C$_{24}$I$_2$   \\
     \hline
  $d$,\AA                 & 3.6      & 6.0          &  7.0           & 7.5            & 7.4            \\
 $\Delta E_{abs}$, meV/mol    & 0      & 648       &  605       & 542         & 114         \\
 $\Delta E_{bind}$, meV/C & -49.5      & -12.9          &  -13.6           & -14.6          & -22.9          \\
    \end{tabular}
\end{ruledtabular}
    \end{table}

It was recently demonstrated that graphene sheets are impermeable
even to light gases including helium.\cite{McEuen,Peeters} For
this reason, we cannot expect that the intercalation of halogen
molecules in graphite is a result of a penetration of the
molecules underneath the top surface layer. On the contrary, we
suppose that the formation of intercalated graphite involves a
consecutive exfoliation of the whole graphite structure.

In Table \ref{grt_table} we summarize equilibrium distances,
absorption energies, and interlayer binding energies calculated
for the investigated systems.\cite{abs-bind} The interlayer
distance in intercalated graphite is approximately twice larger
than in the pure case. We note, however, that the estimation of
the interlayer spacing is not very accurate since we ignore in our
calculations a distortion of the intercalated graphite lattice,
which will inevitably take place in the real situation. The
absorption energies given in Table \ref{grt_table} are positive
because their calculation involves a positive energy of the
exfoliation process. It should be pointed out that among the other
considered molecules, iodine molecule has a minimal absorption
energy value.

The binding energies obtained for the pure graphite are in a good
agreement with the previous vdW-DF calculations.\cite{chakarova,Ziambaras} 
Comparing the interlayer
binding for the pure and for the intercalated graphite, we conclude
that the binding becomes significantly weaker in the presence of
halogen molecules. Moreover, the binding energy is lower for the
more massive molecules because of their stronger interaction with
the carbon layers, as discussed in Sec.III. Our results are in a
qualitative agreement with the previous LDA calculations for
bromine-intercalated graphite.\cite{Br-intercalat} Although the
interlayer binding for our particular impurity concentration seems
to be overestimated (in absolute values) in comparison with the
LDA results, the general tendency is reproduced quite well and allows
to confirm the interpretation of the experiment on sonochemical
exfoliation of intercalated graphite given in Ref.\onlinecite{Br-intercalat}.

\section{Conclusion}

We have performed a first-principles investigation of the
adsorption of diatomic halogen molecules on graphene. We have
shown that a major part of the binding energy in the system
corresponds to van der Waals interaction, and, therefore, cannot be
properly described by means of standard (semi-)local
approximations to exchange-correlation energy. It has been found
that even at the large distances (up to 10 \AA) there is a
non-zero interaction between graphene and F$_2$, Br$_2$, I$_2$
molecules. This interactions are of ionic nature and arise due to
the charge transfer from graphene to the molecules.

In contrast to F$_2$ and Cl$_2$ adsorbates, the densities of
states for graphene in the presence of Br$_2$ and I$_2$ molecules
exhibit impurity bands right above the Fermi level, which can
result, for large enough concentration, in the metallization of
graphene. Moreover, the width of the impurity band is directly
dependent on the impurity concentration.

Finally, investigating properties of intercalated graphite we have
found that interlayer binding becomes significantly weaker than in
pure graphite. This is in agreement with the previous experimental
and theoretical investigations of intercalated graphite.

\section{Acknowledgments}
We would like to thank Irina Grigorieva and Tim Wehling for helpful
discussions. We also appreciate the assistance of Claudia Ambrosch-Draxl and 
Dmitrii Nabok for providing a very efficient
algorithm for the calculation of vdW-DF correlation energies by Monte-Carlo
method. The authors acknowledge support from the Cluster of Excellence 
``Nanospintronics'' (Hamburg, Germany) and from Stichting voor Fundamenteel 
Onderzoek der Materie (FOM), the Netherlands.


\begin{thebibliography}{99}

\bibitem{Novoselov}
K.~S. Novoselov, A.~K. Geim, S.~V. Morozov, D.~Jiang, Y.~Zhang, S.~V. Dubonos, I.~V. Grigorieva, and A.~A. Firsov,
Science {\bf 306}, 666 (2004).

\bibitem{Geim}
A.~K. Geim and K.~S. Novoselov,
Nature Materials {\bf 6}, 183 (2007).


\bibitem{Graphene-RMP}
A.~H. Castro Neto, F. Guinea, N.~M.~R. Peres, K.~S. Novoselov, and A.~K. Geim,
Rev. Mod. Phys. {\bf 81}, 109 (2009).

\bibitem{Wehling-PRB}
T.~O. Wehling, M.~I. Katsnelson, and A.~I. Lichtenstein,
Phys. Rev. B. {\bf 80}, 085428 (2009).

\bibitem{Wehling-APL}
T.~O. Wehling, A.~I. Lichtenstein, and M.~I. Katsnelson,
Appl. Phys. Lett. {\bf 93}, 202110 (2009).

\bibitem{Khomyakov}
P.~A. Khomyakov, G. Giovannetti, P.~C. Rusu, G. Brocks, J. van den Brink, and
P.~J. Kelly,
Phys. Rev. B. {\bf 79}, 195425 (2009).

\bibitem{Johll}
H. Johll, H.~C. Kang, and E.~S. Tok,
Phys. Rev. B. {\bf 79}, 245416 (2009).

\bibitem{Br-intercalat}
E. Widenkvist, D.~W. Boukhvalov, S. Rubino, S. Akhtar, J. Lu, R.~A. Quinlan, M.~I. Katsnelson, K. Leifer, H. Grennberg, and U. Jansson,
J. Phys. D: Appl. Phys. {\bf 42}, 112003 (2009).

\bibitem{intercalation-book}
T. Enoki, M. Suzuki, M. Endo,
\emph{Graphite intercalation compounds and applications},
(Oxford University Press, Oxford and New York, 2003).

\bibitem{Dion}
M. Dion, H. Rydberg, E. Schr\"{o}der, D.~C. Langreth, and B.~I. Lundqvist,
Phys. Rev. Lett. {\bf 92}, 246401 (2004).

\bibitem{Thonhauser}
T. Thonhauser, V.~R. Cooper, S. Li, A. Puzder, P. Hyldgaard, and D.~C. Langreth.
Phys. Rev. B {\bf 76}, 125112 (2007).

\bibitem{Baskin}
Y. Baskin and L. Meyer,
Phys. Rev. {\bf 100}, 544 (1955).

\bibitem{Pauling}
L. Pauling, \emph{The nature of the chemical bond and the structure of
molecules and crystals: an introduction to modern structural chemistry}, 3rd
edition (Ithaca, NY: Cornell University Press, 1960).

\bibitem{espresso}
P. Giannozzi, S. Baroni, N. Bonini, M. Calandra, R. Car, C. Cavazzoni, 
D. Ceresoli, G.~L. Chiarotti, M. Cococcioni, I. Dabo, A. Dal Corso, S. de 
Gironcoli, S. Fabris, G. Fratesi, R. Gebauer, U. Gerstmann, C. Gougoussis, 
A. Kokalj, M. Lazzeri, L. Martin-Samos, N. Marzari, F. Mauri, R. Mazzarello, 
S. Paolini, A. Pasquarello, L. Paulatto, C. Sbraccia, S. Scandolo, 
G. Sclauzero, A.P. Seitsonen, A. Smogunov, P. Umari, and R.~M. Wentzcovitch,
J. Phys.: Condens. Matter, {\bf 21}, 395502 (2009).

\bibitem{pseudo}
The pseudopotentials used in this work were taken from 
{\sc Quantum-ESPRESSO} web page http://www.quantum-espresso.org

\bibitem{pbe}
J.~P. Perdew, K. Burke, and M. Ernzerhof,
Phys. Rev. Lett. {\bf 77}, 3865 (1996).

\bibitem{tetrahedron}
P.~E. Bl\"{o}chl, O. Jepsen, and O.~K. Andersen,
Phys. Rev. B. {\bf 49}, 16223 (1994).

\bibitem{monkhorst}
H.~J. Monkhorst and J.~D. Pack,
Phys. Rev. B. {\bf 13}, 5188 (1976).

\bibitem{Andersson}
Y. Andersson, D.~C. Langreth, and B.~I. Lundqvist,
Phys. Rev. Lett. {\bf 76}, 102 (1996).

\bibitem{Silvestrelli-JPCA}
P.~L. Silvestrelli
J. Phys. Chem. A {\bf 113}, 5224 (2009).

\bibitem{Nguyen}
H.-V. Nguyen and S. de Gironcoli,
Phys. Rev. B {\bf 79}, 115105 (2009).

\bibitem{Niquet}
Y.~M. Niquet, M. Fuchs, and X. Gonze,
Phys. Rev. A {\bf 68}, 032507 (2003).

\bibitem{Harl}
J. Harl and G. Kresse,
Phys. Rev. B {\bf 77}, 045136 (2008).

\bibitem{Langreth}
D.~C. Langreth, B.~I. Lundqvist, S.~D. Chakarova-K\"{a}ck, V.~R. Cooper, M. Dion, P. Hyldgaard, A. Kelkkannen, J. Kleis, Lingzhu Kong,
Shen Li, P.~G. Moses, E. Murray, A. Puzder, H. Rydberg, E. Schr\"{o}der, and T. Thonhauser,
J. Phys.: Condens. Matter {\bf 21}, 084203 (2009).

\bibitem{Zhang-Yang}
Y. Zhang and W. Yang,
Phys. Rev. Lett. {\bf 80}, 890 (1998).

\bibitem{gross}
A. Gross, \emph{Theoretical surface science: a microscopic perspective}
(Springer-Verlag Berlin Heidelberg, 2003), p.171.

\bibitem{grigorieva}
I.~V. Grigorieva, private communication.

\bibitem{comment}
We calculate the charge transfer following the procedure used in Ref.\onlinecite{Khomyakov}. This procedure implies that the
charge $q$ can be estimated by integrating plane-averaged density difference $\Delta n(z)$ as follows:
$q=-|e|\int_{z_0}^{\infty}dz \Delta n(z)$, where $z_0$ corresponds to an interface point so that $\Delta n(z_0)=0$.

\bibitem{Grt-vdw1}
J.-C. Charlier, X. Gonze, and J.-P. Michenaud,
Europhys. Lett. {\bf 28}, 403 (1994).

\bibitem{Grt-vdw2}
L.~A. Girifalco and M. Hodak,
Phys. Rev. B {\bf 65}, 125404 (2002).

\bibitem{Grt-vdw3}
Y.~J. Dappe, M.~A. Basanta, F. Flores, and J. Ortega,
Phys. Rev. B {\bf 74}, 205434 (2006).

\bibitem{PCCP}
A.~H.~R. Palser,
Phys. Chem. Chem. Phys. {\bf 1}, 4459 (1999).

\bibitem{chakarova}
S.~D. Chakarova-K\"{a}ck, E. Schr\"{o}der, B.~I. Lundqvist, and D.~C. Langreth,
Phys. Rev. Lett. {\bf 96}, 146107 (2006).

\bibitem{benedict}
L.~X. Benedict, N.~G. Chopra, M.~L. Cohen, A. Zettl, S.~G. Louie, and V.~H. Crespi,
Chem. Phys. Lett. {\bf 286}, 490 (1998).

\bibitem{zacharia}
R. Zacharia, H. Ulbricht, and T. Hertel,
Phys. Rev. B {\bf 69}, 155406 (2004).

\bibitem{Ziambaras}
E. Ziambaras, J. Kleis, E. Schr\"{o}der, and P. Hyldgaard,
Phys. Rev. B {\bf 76}, 155425 (2007).

\bibitem{McEuen}
J.~S. Bunch, S.~S. Verbridge, J.~S. Alden, A.~M. van der Zande, J.~M. Parpia,
H.~G. Craighead, and P.~L. McEuen,
Nano Lett. {\bf 8}, 2458 (2008).

\bibitem{Peeters}
O. Leenaerts, B. Partoens, and F.~M. Peeters,
Appl. Phys. Lett. {\bf 93}, 193107 (2008).

\bibitem{abs-bind}
We calculate the absorption energy of intercalated graphite as follows:
$\Delta E_{abs}=-\Delta E_{form}^{C_{48}} + \Delta E_{form}^{C_{48}X_4}$, where the first term corresponds to the energy of the pure
graphite formation from a number of non-interacting graphene sheets, and the seconds term is the energy of the intercalated graphite
formation. In turn, the binding energy of intercalated graphite we estimate as:
$\Delta E_{bind}=\Delta E_{form}^{C_{48}X_4}-2\Delta E_{form}^{C_{24}X_2}$, where second term is the binding energy in graphene-halogen
system. X$_2$=F$_2$,Cl$_2$,Br$_2$,I$_2$.


\end{thebibliography}
\end{document}